\documentclass[aps,pra,reprint,superscriptaddress,showpacs,floatfix,preprintnumbers,showkeys]{revtex4-1}

\usepackage{graphicx}
\usepackage{color}
\usepackage{mathtools}
\usepackage{dcolumn}
\usepackage[american]{babel}
\usepackage[colorlinks,breaklinks=true,urlcolor=blue,citecolor=blue,linkcolor=blue,pdfstartview=FitH,pdfpagemode=UseNone]{hyperref}
\usepackage{siunitx}
\usepackage[utf8]{inputenc}

\usepackage[T1]{fontenc}
\usepackage{txfonts}

\mathtoolsset{showonlyrefs=false}

\DeclareSIUnit{\atomicpercent}{at.\%}

\newcommand{\abs}[1]{\lvert #1 \rvert}                      
\newcommand{\elemindex}[1]{{\nu_{#1}}}
\newcommand{\inda}{i} 
\newcommand{\indb}{j} 
\newcommand{\indc}{k} 
\newcommand{\VZBL}[2]{V^{\mathrm{ZBL}}_{{#1}{#2}}}
\newcommand{\VABOP}[2]{V^{\mathrm{Ter}}_{{#1}{#2}}}
\newcommand{\target}{T}
\newcommand{\Ndata}{N_{\mathrm{data}}}
\newcommand{\targetvalue}{t}
\newcommand{\prediction}{p}
\newcommand{\weight}{w}

\newcommand{\eseg}[1]{E_{\mathrm{segr}}^{\mathrm{L}{#1}}}
\newcommand{\etot}{E_{\mathrm{tot}}}
\newcommand{\ecoh}{E_{\mathrm{coh}}}
\newcommand{\eform}{E_{\mathrm{f}}}
\newcommand{\Fe}{\mathrm{Fe}}
\newcommand{\Cr}{\mathrm{Cr}}

\newcommand{\Crconc}{c_{\mathrm{Cr}}}

\newcommand{\softpekcom}[1]{}
\newcommand{\akcom}[1]{\textcolor{green}{$\langle$AK: #1$\rangle$}}

\begin{document}
\title{Interatomic Fe--Cr potential for modeling kinetics on Fe surfaces}
\date{\today}

\author{Pekko Kuopanportti}\email{pekko.kuopanportti@gmail.com}
\affiliation{Department of Physics, University of Helsinki, P.O. Box 43, FI-00014 Helsinki, Finland}
\author{Matti Ropo}
\affiliation{Department of Physics and Astronomy, University of Turku, FI-20014 Turku, Finland}
\author{Daniel Holmberg}
\affiliation{Department of Physics, University of Helsinki, P.O. Box 43, FI-00014 Helsinki, Finland}
\author{Henrik Levämäki}
\affiliation{Department of Physics and Astronomy, University of Turku, FI-20014 Turku, Finland}
\author{Kalevi Kokko}
\affiliation{Department of Physics and Astronomy, University of Turku, FI-20014 Turku, Finland}
\author{Sari Granroth}
\affiliation{Department of Physics and Astronomy, University of Turku, FI-20014 Turku, Finland}
\author{Antti Kuronen}
\affiliation{Department of Physics, University of Helsinki, P.O. Box 43, FI-00014 Helsinki, Finland}


\begin{abstract}
To enable accurate molecular dynamics simulations of iron--chromium alloys with surfaces, we develop, based on density-functional-theory (DFT) calculations, a new interatomic Fe--Cr potential in the Tersoff formalism. Contrary to previous potential models, which have been designed for bulk Fe--Cr, we extend our potential fitting database to include 
not only conventional bulk properties but also surface-segregation energies of Cr in bcc Fe.
In terms of reproducing our DFT results for the bulk properties, the new potential is found to be superior to the previously developed Tersoff potential and competitive with the concentration-dependent and two-band embedded-atom-method potentials.
For Cr segregation toward the surface of an Fe--Cr alloy, 
only the new potential agrees with our DFT calculations in predicting preferential segregation of Cr to the topmost surface layer, instead of the second layer preferred by the other potentials.
%
%
%
%
%
%
We expect this rectification 
to foster future research, e.g., on
the mechanisms of corrosion resistance of stainless steels at the atomic level.
\end{abstract}
\keywords{Iron--chromium alloy, Molecular dynamics, Tersoff potential}

\maketitle




\section{Introduction}


Iron--chromium alloys are not only scientifically interesting due to their peculiar material properties, but also play an important technological role as the base component for stainless steels \cite{Vit2011.chapter,Wen2011.book}. 
From a basic-science perspective,
Fe--Cr alloys at varying relative compositions and structures exhibit intricate phenomena such as giant magnetoresistance~\cite{Gru1986.PRL57.2442}, a spin-ice phase~\cite{Bur1983.JPhysF13.451}, and the ``\SI{475}{\celsius} embrittlement'' effect~\cite{Sah2009.MaterSciEngA508.1}. 
From an application point of view,
precision design of steels---beyond the traditional method of empirical trial and error---benefits from an atomic-level understanding of the ins and outs of Fe--Cr alloys.
Of particular importance is to explore the mechanisms by which the Cr atoms render the open surfaces of Cr-containing steels 
resistant to corrosion 
at and above Cr concentrations of 9--10\%~\cite{Lev1989.Wear131.39}.



The static properties of the iron--chromium system have been extensively studied at the atomic level. For example, surface~\cite{Rop2011.JPCM23.265004} and interface~\cite{Lu2011.PhysStatusSolidiB248.2087} energies and surface segregation energies of chromium~\cite{Lev2012.PRB85.064111} have been calculated for Fe--Cr alloys using \emph{ab initio} methods.
Importantly, such calculations have shown that the aforementioned critical Cr concentration for the onset of the corrosion resistance in stainless steels closely coincides with an onset of anomalous surface segregation of Cr in Fe--Cr alloys~\cite{Rop2007.PRB76.220401, Lev2013.PRB87.075409}. This observation suggests that the segregation of Cr toward the surface plays a crucial role in facilitating the formation of the protective, self-healing layer of iron and chromium oxides that is known to be the reason for the corrosion resistivity of stainless steels~\cite{Ole1980.MaterSciEng42.161,Lin1992.SurfSci277.43,Gon2007.InorgMater43.515,Don2009.CorrosSci51.827,Rop2021.SciRep11.6046}. 


The current \emph{ab initio} modeling methods give reasonably accurate predictions of the static properties of Fe--Cr alloys. The drawback of these methods is that they are computationally heavy, 
to the extent that the length scales in the modeling can be no larger than a few nanometers and that the costs of studying time-dependent phenomena are still largely prohibitive.
In an attempt to overcome these limitations, semi-empirical potential models have been developed for the Fe--Cr system.
The most recent of these 
are the concentration-dependent embedded-atom model~(CDEAM)~\cite{Car2005.PRL95.075702,*Stu2009.ModelSimulMaterSc17.075005}, the two-band embedded-atom model~(2BEAM)~\cite{Bon2011.PhilosMag91.1724}, and 
the Tersoff potential~\cite{Hen2013.JPCM25.445401}. 
These potentials have been fitted to various basic material properties such as cohesion energies, lattice constants, and elastic properties, and, in general, they describe the point-defect energetics and the solubility of chromium in iron at low concentrations fairly well.
As such, the semi-empirical Fe--Cr potential models are well suited for modeling bulk Fe--Cr alloys and have been employed, in conjunction with Monte Carlo methods, to study 
diffusion coefficients and precipitation kinetics~\cite{Sen2014.ActaMater73.97}, vacancy migration near grain boundaries~\cite{Cas2014.ComputMaterSci84.217}, and equilibrium configurations of phase-separated Fe--Cr alloys~\cite{Zhu2011.JNuclMater417.1082} (using an earlier 2BEAM parametrization~\cite{Ols2005.PRB72.214119}).


The aforementioned semi-empirical potential models~\cite{Bon2011.PhilosMag91.1724,Car2005.PRL95.075702,Stu2009.ModelSimulMaterSc17.075005,Hen2013.JPCM25.445401} turn out to be, however, much less successful in predicting
properties of Fe--Cr surfaces~\cite{Kur2015.PRB92.214113}. 
Their disagreement with 
\emph{ab initio} calculations is succinctly demonstrated by inspecting the segregation of a Cr atom from the bulk to the surface of a bcc Fe crystal: all three semi-empirical models predict the Cr atom to segregate to the second layer of Fe atoms, whereas according to \emph{ab initio} calculations the first layer (i.e., the surface itself) should be preferred~\cite{Kur2015.PRB92.214113}. In addition, all three models fail to reproduce the \emph{ab initio} results of Fe--Cr interface energies~\cite{Lu2011.PhysStatusSolidiB248.2087}. As a consequence of these shortcomings, no Monte Carlo simulations of atom kinetics 
near Fe--Cr surfaces
have been performed, and the current atomic-level knowledge of surface physics in Fe--Cr alloys is far from satisfactory, 
despite the practical importance of the topic in relation to the
corrosion of steel surfaces.




In this paper, we develop a new semi-empirical potential model that is suitable for investigating surface physics in Fe--Cr alloys at the atomic level,
yet offers performance on par with the existing models in describing the bulk alloy.
%
%
%
%
We choose to work in the Tersoff formalism and use previously developed potentials for the homonuclear Fe--Fe~\cite{Mul2007.JPCM19.326220,Bjo2007.NIMB259.853} and Cr--Cr~\cite{Hen2013.JPCM25.445401} interactions while constructing a new one for the heteronuclear Fe--Cr interaction.
%
This choice 
of homonuclear potentials 
makes the new potential model directly compatible with the previously developed Fe--C~\cite{Hen2009.PRB79.144107,Hen2013.JPCM25.445401} and Cr--C~\cite{Hen2013.JPCM25.445401} models~\footnote{The Fe--C and Cr--C models both use the same C--C potential from Refs.~\cite{Bre1990.PRB42.9458,Jus2005.JApplPhys98.123520}, and are therefore mutually compatible.}, so that they can all be combined into an Fe--Cr--C potential model for stainless steel.  




The remainder of this paper is organized as follows. In Sec.~\ref{sec:methods}, we 
introduce 
the Tersoff potential formalism, describe the fitting procedure by which we develop the new Fe--Cr potential, and 
outline 
the DFT methods 
that we use to generate the target data for the fitting procedure.
Section~\ref{sec:results} is devoted to presenting the new potential and benchmarking it against the pre-existing potential models and DFT calculations, 
with both bulk and surface properties considered. 
\softpekcom{Removed a false sentence about the contents. The results mentioned could possibly be added into the paper, though.}
Finally, we summarize our main results and discuss their implications in Sec.~\ref{sec:discussion}.

\section{Theory and methods}\label{sec:methods}

\subsection{Potential formalism}\label{subsec:potential_formalism}

The reactive Tersoff formalism~\cite{Ter1988.PRB37.6991,Bre1990.PRB42.9458,Alb2002.PRB65.195124} adopted in this work originates from Pauling's concept of bond order~\cite{Pau1960.book,Abe1985.PRB31.6184};
it can also be formally linked~\cite{Alb2002.PRB65.195124} to
both the tight-binding scheme~\cite{Cle93.PRB48.22} and the embedded-atom method~\cite{Daw1984.PRB29.6443,Bre1989.PRL63.1022}. Since the same formalism has already been described extensively elsewhere~\cite{Alb2002.PRB65.195124,Alb2002.PRB66.035205,Nor2003.JPCM15.5649,Jus2005.JApplPhys98.123520}, we will give here only a brief overview.
In the molecular dynamics code {\footnotesize{LAMMPS}}~\cite{Pli1995.JCompPhys117.1}, this formalism is available as the potential style \texttt{tersoff/zbl}.
Although the only atomic types considered in this work are the elements Fe and Cr, we present the potential formalism below for a general system with an arbitrary number of atomic types.

Let each atom in the system (regardless of its type) be assigned a unique ordinal number, which we will denote using the roman indices $\inda,\indb,\indc\in\mathbb{N}$. Furthermore, let $\left(\elemindex{\inda}\right)_{\inda}$ be a finite sequence such that $\elemindex{\inda}$ gives the type of the $\inda$th atom, with $\inda \in \left\{1,\dots,N\right\}$ and $N$ denoting the total number of atoms in the system~\footnote{In our case, $\elemindex{\inda}\in \left\{ \mathrm{Fe},\mathrm{Cr} \right\}\ \forall \inda\left\{1,\dots,N\right\}$ and $N=N_\Fe+ N_\Cr$.}. 
In the Tersoff formalism, the total potential energy $E_\mathrm{tot}$ of the system can then be written as a sum of individual bond energies:
\begin{equation}\label{eq:total_potential}
E_\mathrm{tot}=\sum_{\inda=1}^N \sum_{\indb = \inda+1}^N \left\{ \VZBL{\elemindex{\inda}}{\elemindex{\indb}} \bigl(r_{\inda\indb}\bigr)\left[1-F_{\elemindex{\inda}\elemindex{\indb}} \bigl(r_{\inda\indb}\bigr)\right]+\VABOP{\elemindex{\inda}}{\elemindex{\indb}}\bigl(r_{\inda\indb}\bigr)F_{\elemindex{\inda}\elemindex{\indb}}\bigl(r_{\inda\indb}\bigr)\right\},
\end{equation}
where $r_{\inda\indb}$ is the distance between atoms $\inda$ and $\indb$, $\VZBL{\elemindex{\inda}}{\elemindex{\indb}}$ is the universal Ziegler--Biersack--Littmark (ZBL) potential~\cite{Zie1985_book,*Zie1985.incollection} for elements $\elemindex{\inda}$ and $\elemindex{\indb}$, \(\VABOP{\elemindex{\inda}}{\elemindex{\indb}}\) is the pure Tersoff potential for these elements, 
and
\begin{equation}\label{eq:fermi}
F_{\elemindex{\inda}\elemindex{\indb}}\bigl(r_{\inda\indb}\bigr)=\left\{1+\exp\left[-b_{\mathrm{F},\elemindex{\inda}\elemindex{\indb}}\left(r_{\inda\indb}-r_{\mathrm{F},\elemindex{\inda}\elemindex{\indb}}\right)\right]\right\}^{-1}
\end{equation}
is a Fermi function used to join the short-range ZBL and longer-range Tersoff parts smoothly together.  The values of the parameters $b_\mathrm{F}$ and $r_\mathrm{F}$ are chosen manually such that the potential is essentially the unmodified Tersoff potential at and past equilibrium bonding distances and that a smooth transition to the ZBL potential at short separations is obtained for all realistic coordination numbers. \softpekcom{Refer to other papers utilizing this same interpolation approach?}

Incorporation of the ZBL potential [as done in Eq.~\eqref{eq:total_potential}] is needed to make the potential formalism suitable for modeling nonequilibrium phenomena such as melting or high-energy particle irradiation processes, which typically involve repulsive short-distance interactions originating mainly from the screened Coulomb repulsion between the positively charged nuclei. The ZBL potential is written as
\begin{equation}\label{eq:ZBL}
\VZBL{\elemindex{\inda}}{\elemindex{\indb}}\bigl(r_{\inda\indb}\bigr)=\frac{e^2}{4\pi \varepsilon_0}\frac{Z_{\elemindex{\inda}}Z_{\elemindex{\indb}}}{r_{\inda\indb}} \phi\bigl(r_{\inda\indb}/a_{\elemindex{\inda}\elemindex{\indb}} \bigr),
\end{equation}
where $Z_\elemindex{\inda}$ is the atomic number of element $\elemindex{\inda}$,
\begin{equation}
a_{\elemindex{\inda}\elemindex{\indb}}= \frac{\SI{0.8854}{\bohr}}{Z_{\elemindex{\inda}}^{0.23}+Z_{\elemindex{\indb}}^{0.23}}
\end{equation}
with $\si{\bohr}$ denoting the Bohr radius, and $\phi$ is the universal screening function
\begin{equation}
\begin{split}
\phi\left(x\right)&=0.1818\,e^{-3.2\, x}+ 0.5099\, e^{-0.9423\,x} \\ & \quad + 0.2802\,  e^{-0.4028\,x} + 0.02817\,e^{-0.2016\,x}.
\end{split}
\end{equation}
This screening function has been fitted to the interaction energy between ions, and its accuracy is $\sim\mkern-6mu 10\%$~\cite{Zie1985_book,*Zie1985.incollection}.

The Tersoff part $\VABOP{}{}$ is what chiefly determines the equilibrium properties of the system. It is written as
\begin{equation}\label{eq:ABOP}
\VABOP{\elemindex{\inda}}{\elemindex{\indb}}\bigl(r_{\inda\indb}\bigr) = f^\mathrm{c}_{\elemindex{\inda}\elemindex{\indb}} \bigl(r_{\inda\indb}\bigr)\left[V_{\elemindex{\inda}\elemindex{\indb}}^\mathrm{R}\bigl(r_{\inda\indb}\bigr) - \frac{b_{\elemindex{\inda}\elemindex{\indb}}+b_{\elemindex{\indb}\elemindex{\inda}}}{2} V_{\elemindex{\inda}\elemindex{\indb}}^\mathrm{A}\bigl(r_{\inda\indb}\bigr) \right],
\end{equation} 
where $f^\mathrm{c}$ is a cutoff function for the pair interaction, $V^\mathrm{R}$ is a repulsive and $V^\mathrm{A}$ an attractive pair potential, and $b$ is a bond-order term that describes three-body interactions and angularity. The pair potentials are of the Morse-like form
\begin{subequations}\label{eq:pair_potentials}
\begin{align}
\label{eq:pair_potentialsR}
V^\mathrm{R}_{\elemindex{\inda}\elemindex{\indb}}\bigl(r_{\inda\indb}\bigr) &=\frac{D_{0,\elemindex{\inda}\elemindex{\indb}}}{S_{\elemindex{\inda}\elemindex{\indb}}-1} \exp\left[- \sqrt{2S_{\elemindex{\inda}\elemindex{\indb}}}\beta_{\elemindex{\inda}\elemindex{\indb}}\left(r_{\inda\indb}-r_{0,\elemindex{\inda}\elemindex{\indb}}\right)\right], \\ 
\label{eq:pair_potentialsA}
V^\mathrm{A}_{\elemindex{\inda}\elemindex{\indb}}\bigl(r_{\inda\indb}\bigr) &=\frac{S_{\elemindex{\inda}\elemindex{\indb}}D_{0,\elemindex{\inda}\elemindex{\indb}}}{S_{\elemindex{\inda}\elemindex{\indb}}-1} \exp\left[-\frac{\sqrt{2}\beta_{\elemindex{\inda}\elemindex{\indb}}}{\sqrt{S_{\elemindex{\inda}\elemindex{\indb}}}}\left(r_{\inda\indb}-r_{0,\elemindex{\inda}\elemindex{\indb}}\right)\right],
\end{align}
\end{subequations}
where $D_0$ and $r_0$ are the bond energy and length of the dimer molecule, respectively, and $S>1$ is 
a dimensionless parameter that adjusts the relative strengths of the repulsive and attractive terms. 
The parameter $\beta$ is related to the ground-state vibrational frequency $\omega$ and the reduced mass $\mu$ of the dimer according to 
\begin{equation} 
\beta_{\elemindex{\inda}\elemindex{\indb}}=\frac{\sqrt{2\mu_{\elemindex{\inda}\elemindex{\indb}}}\pi\omega_{\elemindex{\inda}\elemindex{\indb}}}{\sqrt{D_{0,\elemindex{\inda}\elemindex{\indb}}}}.
\end{equation}
The bond-order term is given by 
\begin{equation}
b_{\elemindex{\inda}\elemindex{\indb}} = \frac{1}{\sqrt{1+\chi_{\elemindex{\inda}\elemindex{\indb}}}},
\end{equation}
where 
\begin{equation}\label{eq:chi}
\chi_{\elemindex{\inda}\elemindex{\indb}}=\sum_{\indc\left(\neq \inda,\indb\right)} f^\mathrm{c}_{\elemindex{\inda}\elemindex{\indc}}\bigl(r_{\inda\indc}\bigr) g_{\elemindex{\inda}\elemindex{\indc}}\bigl(\theta_{\inda\indb\indc}\bigr)\exp\bigl[\alpha_{\elemindex{\inda}\elemindex{\indb}\elemindex{\indc}}\bigl(r_{\inda\indb}-r_{\inda\indc} \bigr)\bigr].
\end{equation} 
In Eq.~\eqref{eq:chi}, $\theta_{\inda\indb\indc}$ is the angle between the vectors $\mathbf{r}_{\inda\indb}= \mathbf{r}_{\indb} - \mathbf{r}_{\inda}$ and $\mathbf{r}_{\inda\indc}$, and $g$ is the angular function 
\begin{equation}\label{eq:g}
g_{\elemindex{\inda}\elemindex{\indc}}\bigl(\theta_{\inda\indb\indc}\bigr)=\gamma_{\elemindex{\inda}\elemindex{\indc}}\left[1+\frac{c_{\elemindex{\inda}\elemindex{\indc}}^2}{d_{\elemindex{\inda}\elemindex{\indc}}^2}-\frac{c_{\elemindex{\inda}\elemindex{\indc}}^2}{d_{\elemindex{\inda}\elemindex{\indc}}^2+\left(h_{\elemindex{\inda}\elemindex{\indc}}+\cos\theta_{\inda\indb\indc}\right)^2}\right],
\end{equation}
where $\gamma$, $c$, $d$, and $h$ are adjustable parameters. 

The cutoff function $f^\mathrm{c}$ appearing in Eqs.~\eqref{eq:ABOP} and~\eqref{eq:chi} is continuously differentiable and is defined piecewise as
\begin{equation}\label{eq:cutoff}
f^\mathrm{c}_{\elemindex{\inda}\elemindex{\indb}}\bigl(r_{\inda\indb}\bigr) =
\begin{cases}
1,&  r_{\inda\indb} < R_{\elemindex{\inda}\elemindex{\indb}}-D_{\elemindex{\inda}\elemindex{\indb}}, \\
\frac{1}{2}-\frac{1}{2}\sin\frac{\pi\left(r_{\inda\indb}-R_{\elemindex{\inda}\elemindex{\indb}}\right)}{2D_{\elemindex{\inda}\elemindex{\indb}}},& |r_{\inda\indb}-R_{\elemindex{\inda}\elemindex{\indb}}|\leq D_{\elemindex{\inda}\elemindex{\indb}}, \\
0,& r_{\inda\indb} > R_{\elemindex{\inda}\elemindex{\indb}}+D_{\elemindex{\inda}\elemindex{\indb}},
\end{cases}
\end{equation} 
where $R$ and $D$ determine, respectively, the center and width of the cutoff interval. Typically, $R$ is chosen to lie midway between the second- and third-nearest neighbors in the relevant equilibrium crystal.

\subsection{Fitting procedure}\label{subsec:fitting}

In order to devise a well-performing Fe--Cr potential in the Tersoff formalism, we use the following approach: 
The parameters for the Cr--Cr interaction are all taken completely unchanged from Ref.~\cite{Hen2013.JPCM25.445401}, while the parameters for the Fe--Fe interaction are taken unchanged from Refs.~\cite{Mul2007.JPCM19.326220,Bjo2007.NIMB259.853} with the exception that the value of the cutoff distance $R_{\mathrm{Fe}\,\mathrm{Fe}}$ is increased to \SI{3.5}{\angstrom} from the original \SI{3.15}{\angstrom} to avoid an unphysical increase in Young's modulus of elasticity at elevated temperatures~\cite{[{This same cutoff adjustment has been previously made by, e.g., }]Kuo2016.CMS111.525}. 
Furthermore, the two parameters $b_{\mathrm{F},\,\Fe\,\Cr}$ and $r_{\mathrm{F},\,\Fe\,\Cr}$ appearing in the Fermi function [Eq.~\eqref{eq:fermi}] of the Fe--Cr potential are chosen to be the same as in Ref.~\cite{Hen2013.JPCM25.445401}. 
After making these initial choices, we are left with 
a total of 17 Fe--Cr potential parameters 
that we then determine by numerical optimization.

%
%

We implement the numerical optimization in {\footnotesize MATLAB}~\cite{matlab}. To this end, the fitting of the 17 non-predetermined potential parameters is formulated as the nonlinear constrained least-squares minimization problem
\begin{equation}\label{eq:minimization_problem}
\min_{\xi_i\in[0,1]}\target(\xi_1,\dots,\xi_{17}),
\end{equation}
where $\xi_i\in[0,1]$  is a normalized optimization variable that maps to the closed interval between the minimum and maximum values we allow for the $i$th potential parameter. 
The minimum (maximum) allowed values are chosen small (large) enough to have no significant impact on the optimal solution.
The target function $\target$ is the weighted square sum of the differences between the target values $\targetvalue_j$ and the potential model's predictions $\prediction_j$,
\begin{equation}\label{eq:target_function}
\target(\xi_1,\dots,\xi_{17})=\sum_{j=1}^{\Ndata}\weight_j \left[ \targetvalue_j - \prediction_j\left(\xi_1,\dots,\xi_{17}\right)\right]^2,
\end{equation}
where $\weight_j \geq 0$ and $\Ndata$ is the number of evaluated quantities in the fitting database. The constrained minimization problem \eqref{eq:minimization_problem} is solved using the trust-region-reflective algorithm~\cite{Mor1983.SJSSC3.553,Byr1988.MaPr40.247,Bra1999.SJSC21.1} implemented in the {\footnotesize MATLAB} function \texttt{lsqnonlin}. The potential predictions $\prediction_j$ are evaluated by calling the \texttt{minimize} command in {\footnotesize LAMMPS}~\cite{Pli1995.JCompPhys117.1}.
The target values in our fitting database are determined beforehand by DFT calculations, with the database consisting of the following quantities ($\Ndata=40$):
\begin{itemize}
    \item FeCr dimer bond length $r_0$ (target value \SI{2.1428}{\angstrom});
    \item formation energies of nine Cr point defects in bcc Fe (see Table~\ref{table:defect_energies} for the specific defects and the target formation energies); 
    \item mixing energy of Fe\textsubscript{52}Cr\textsubscript{2} (7 different configurations);
    \item mixing energy of Fe\textsubscript{51}Cr\textsubscript{3} (16 different configurations);
    \item mixing energy of Fe\textsubscript{40}Cr\textsubscript{14} (SQS cell);
    \item mixing energy of Fe\textsubscript{27}Cr\textsubscript{27} (SQS cell);
    \item mixing energy of Fe\textsubscript{14}Cr\textsubscript{40} (SQS cell);
    \item Cr segregation energies to the first four layers of a (100) surface of bcc Fe.
\end{itemize}
Here the alloy mixing energies are calculated for a periodically repeating 54-atom bcc 
cell; the subscripts indicate the numbers of Fe and Cr atoms per cell. 
For computational efficiency, the mixing energies corresponding to the combinatorially challenging intermediate Cr concentrations from 25.9\% to 74.1\% are estimated by means of special quasirandom structures (SQSs)~\cite{Zun1990.PRL65.353}. For a given Cr concentration $\Crconc \coloneqq N_\Cr/\left(N_\Fe+N_\Cr\right)$, the SQS cell is a single, computationally constructed 54-atom cell whose lattice sites are occupied by Fe and Cr atoms in such a way that the resulting periodic lattice mimics as closely as possible the first few, physically most relevant radial correlation functions of a perfectly random lattice with the same $\Crconc$. We generate the SQS cells using the \texttt{mcsqs} code~\cite{Wal2013.Calphad42.13} of the Alloy Theoretic Automated Toolkit~\cite{Wal2002.Calphad26.539}.


\subsection{DFT calculations}\label{subsec:dft_methods}

Before solving the optimization problem~\eqref{eq:minimization_problem}, we carry out \emph{ab initio} DFT calculations to determine the target values $t_j$ in Eq.~\eqref{eq:target_function}.
All our DFT calculations are performed using the GPAW code~\cite{Mor2005.PRB71.035109,gpaw2} (version 1.1.0) and the Atomic Simulation Environment (ASE)~\cite{ase} (version 3.11). Valence--core interactions are modeled with the projector augmented-wave method (GPAW/PAW version 0.8). 
The generalized-gradient approximation is used in the form of the Perdew--Burke--Ernzerhof exchange–correlation functional~\cite{pbe}.

All the bulk DFT calculations are carried out using a $3\times 3\times 3$ cubic simulation cell of 54 atoms. 
Wave functions are represented on a real-space grid of $48\times 48 \times 48$ points, and Brillouin-zone integrations are performed on 
a Monkhorst--Pack grid~\cite{Mon1976.PRB13.5188} of $4\times 4 \times 4$ $k$ points.
All atomic coordinates are relaxed using the Broyden--Fletcher--Goldfarb--Shanno algorithm 
until the Hellmann--Feynman forces are less than \SI[per-mode=symbol]{0.05}{\electronvolt\per\angstrom}. 

The DFT calculations involving an Fe(100) surface are performed for a $2\times 2\times 5$ slab geometry with a distance of \SI{24}{\angstrom} between the two surfaces. 
A real-space grid of $48\times 48 \times 288$ points and a Monkhorst--Pack grid of $4 \times 4 \times 1$ $k$ points are used. 
The two centermost atomic layers of the slab are fixed to their bulk positions, while all other atoms are relaxed using the {\footnotesize{FIRE}} algorithm~\cite{Bit2006.PRL97.170201} with a force tolerance of \SI[per-mode=symbol]{0.05}{\electronvolt\per\angstrom}.

\section{Results}\label{sec:results}

\subsection{New Fe--Cr potential}

\softpekcom{If needed, discuss the values of the cutoff parameters here.}

The numerically optimized parameter values for the new Fe--Cr Tersoff potential are shown in Tables~\ref{table:abop_parameters} and~\ref{table:alpha_parameters}. The corresponding potential input file for the LAMMPS potential style \texttt{tersoff/zbl} is available at [URL].

It is worth noting from Table~\ref{table:alpha_parameters} that the values of the six three-body parameters $\alpha$ pertaining to the heteronuclear Fe--Cr part have apparently not been numerically optimized in the old Tersoff parametrization of Ref.~\cite{Hen2013.JPCM25.445401}, on the grounds of them all having the same exact value of 1. 
This should be contrasted with 
the new parametrization, where all six heteronuclear $\alpha$ parameters have been fully incorporated into the optimization. 
The resulting increase in the dimensionality of the optimization space may in part explain why we have succeeded in significantly improving the Tersoff potential's agreement with the \emph{ab initio} target data 
for both bulk and surface properties, as demonstrated below.





\begin{table}[htb]
\caption{\label{table:abop_parameters}Parameters for the new Tersoff potential of the Fe--Cr system. 
The Fe--Fe part is the same as originally introduced in Ref.~\cite{Mul2007.JPCM19.326220} and subsequently augmented with the additional parameters $b_\mathrm{F}$ and $r_\mathrm{F}$ [Eq.~\eqref{eq:fermi}] in Ref.~\cite{Bjo2007.NIMB259.853}, except for a slightly larger cutoff $R$ that we adopt to avoid an unphysical increase in Young's modulus of elasticity at elevated temperatures; 
the Cr--Cr part is from Ref.~\cite{Hen2013.JPCM25.445401}; the Fe--Cr part given in the last column is derived in this work. 
Note that the three-body parameter $\alpha$ has only one value associated with each of the two homonuclear potentials ($\alpha_{\Fe\Fe\Fe}$ and $\alpha_{\Cr\Cr\Cr}$, as shown here) but assumes six distinct values for the heteronuclear Fe--Cr part (as listed separately in Table~\ref{table:alpha_parameters}). 
All the two-body Fe--Cr parameters  
are symmetric with respect to interchange of the atomic types (i.e., $\gamma_{\mathrm{Fe}\,\mathrm{Cr}}=\gamma_{\mathrm{Cr}\,\mathrm{Fe}}$, and similarly for the others).
}
\begin{ruledtabular}
\begin{tabular}{llddd}
& & \multicolumn{3}{c}{Interaction}\\
\cline{3-5}
\multicolumn{2}{c}{Parameter}&\multicolumn{1}{c}{Fe--Fe} &\multicolumn{1}{c}{Cr--Cr}&\multicolumn{1}{c}{Fe--Cr}\\
\hline
$D_0$ &(eV) & 1.5 & 4.0422 & 1.2277 \\ 
$r_0$ &(\AA) & 2.29 & 2.1301 & 2.2320 \\ 
$\beta$ &(\AA$^{-1}$)& 1.4 & 1.6215 &0.8957\\
$S$ & & 2.0693 & 3.3679 & 3.1743\\  
$\gamma$ & & 0.01158 &  0.1577 & 0.0996 \\ 
$c$ & & 1.2899 & 1.0329 & 0.0794\\ 
$d$ & &  0.3413 & 0.1381 & 5.9464 \\  
$h$ & &  -0.26 & -0.2857 &0.2952 \\
$R$ &(\AA) & 3.5 & 3.2 & 3.15 \\
$D$ &(\AA) & 0.2 & 0.2  & 0.15\\
$\alpha$ & & 0 & 1.3966 & - \\ 
$b_\mathrm{F}$ &($\mathrm{\AA}^{-1}$) & 2.9 & 12.0 & 10.0 \\
$r_\mathrm{F}$ &(\AA) & 0.95 & 1.7 & 1.0
\end{tabular}
\end{ruledtabular}
\end{table}

\begin{table}[htb]
\caption{\label{table:alpha_parameters} Three-body coefficients $\alpha$ pertaining to the Fe--Cr part of the potential for both the old and the new Tersoff parametrizations. The rest of the parameter values for the new Tersoff potential are given in Table~\ref{table:abop_parameters}, 
while the old Tersoff potential is presented in its entirety in Ref.~\cite{Hen2013.JPCM25.445401}.}
\begin{ruledtabular}
\begin{tabular}{ldd}
& \multicolumn{2}{c}{Fe--Cr parametrization}\\
\cline{2-3}
\multicolumn{1}{c}{Parameter}&\multicolumn{1}{c}{Old Tersoff} &\multicolumn{1}{c}{New Tersoff}\\
\hline
$\alpha_{\mathrm{Fe}\,\mathrm{Fe}\,\mathrm{Cr}}$ & 1.0 & -0.356010 \\
$\alpha_{\mathrm{Fe}\,\mathrm{Cr}\,\mathrm{Fe}}$ & 1.0 & -3.282748 \\
$\alpha_{\mathrm{Cr}\,\mathrm{Fe}\,\mathrm{Fe}}$ & 1.0 & -1.181222 \\
$\alpha_{\mathrm{Fe}\,\mathrm{Cr}\,\mathrm{Cr}}$ & 1.0 & -1.340769 \\
$\alpha_{\mathrm{Cr}\,\mathrm{Fe}\,\mathrm{Cr}}$ & 1.0 & -0.233643 \\
$\alpha_{\mathrm{Cr}\,\mathrm{Cr}\,\mathrm{Fe}}$ & 1.0 & -1.455751 
\end{tabular}
\end{ruledtabular}
\end{table}

\subsection{Comparison of the potential models in regard to bulk Fe--Cr}\label{subsec:bulk_results}

\softpekcom{Although it might be beneficial in some sense to start with the surface properties, I think it is conceptually more straightforward to start with the bulk properties first. If you think otherwise, I'm open to changing this. One might also question if the bulk/surface division is the best way to go for us. As for now, there are quite a lot more bulk than surface results---regardless of our project aims.}

\begin{figure}[htb]
\includegraphics[width=0.99\columnwidth,keepaspectratio]{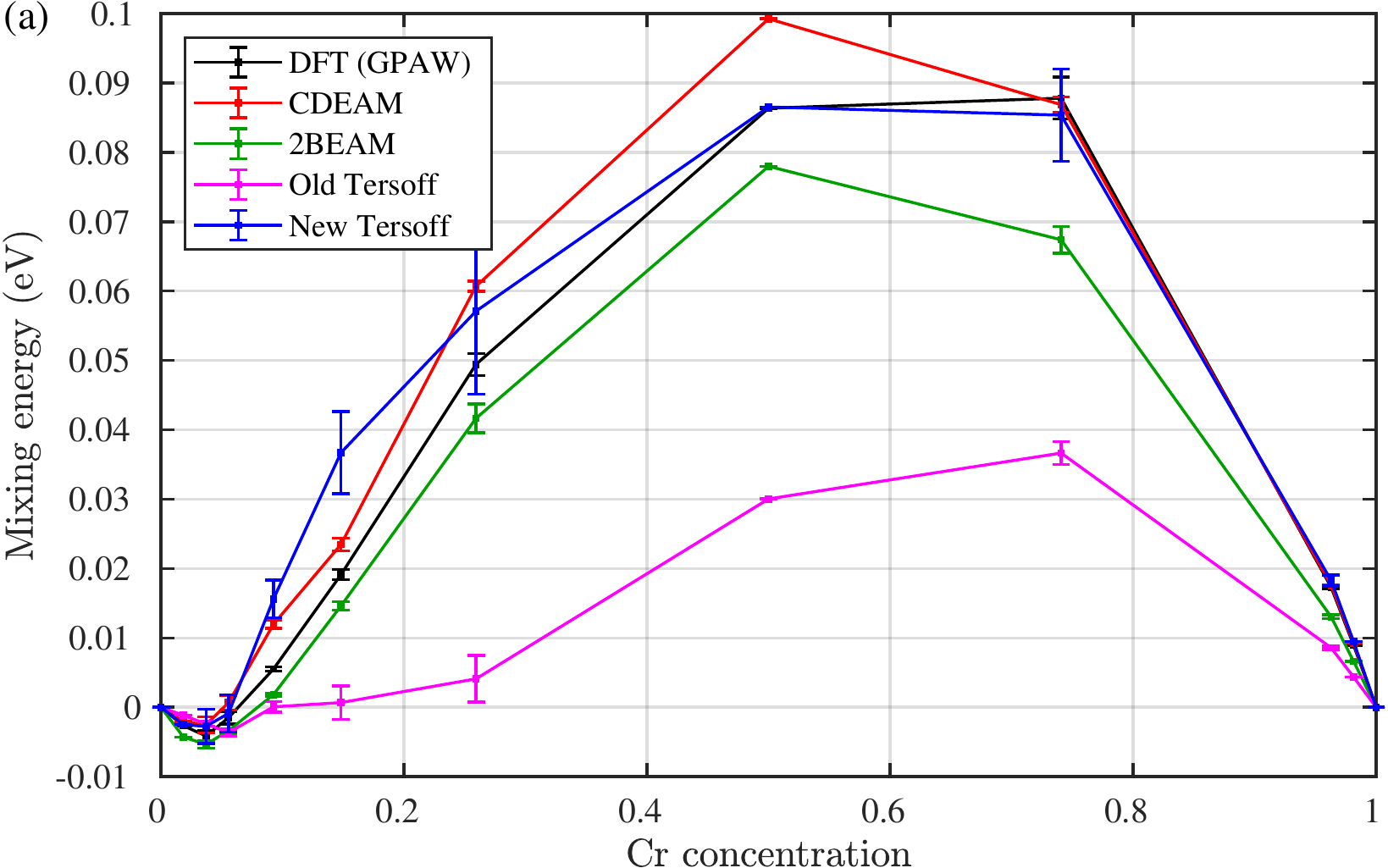}\\
\vspace{2mm}
\includegraphics[width=0.99\columnwidth,keepaspectratio]{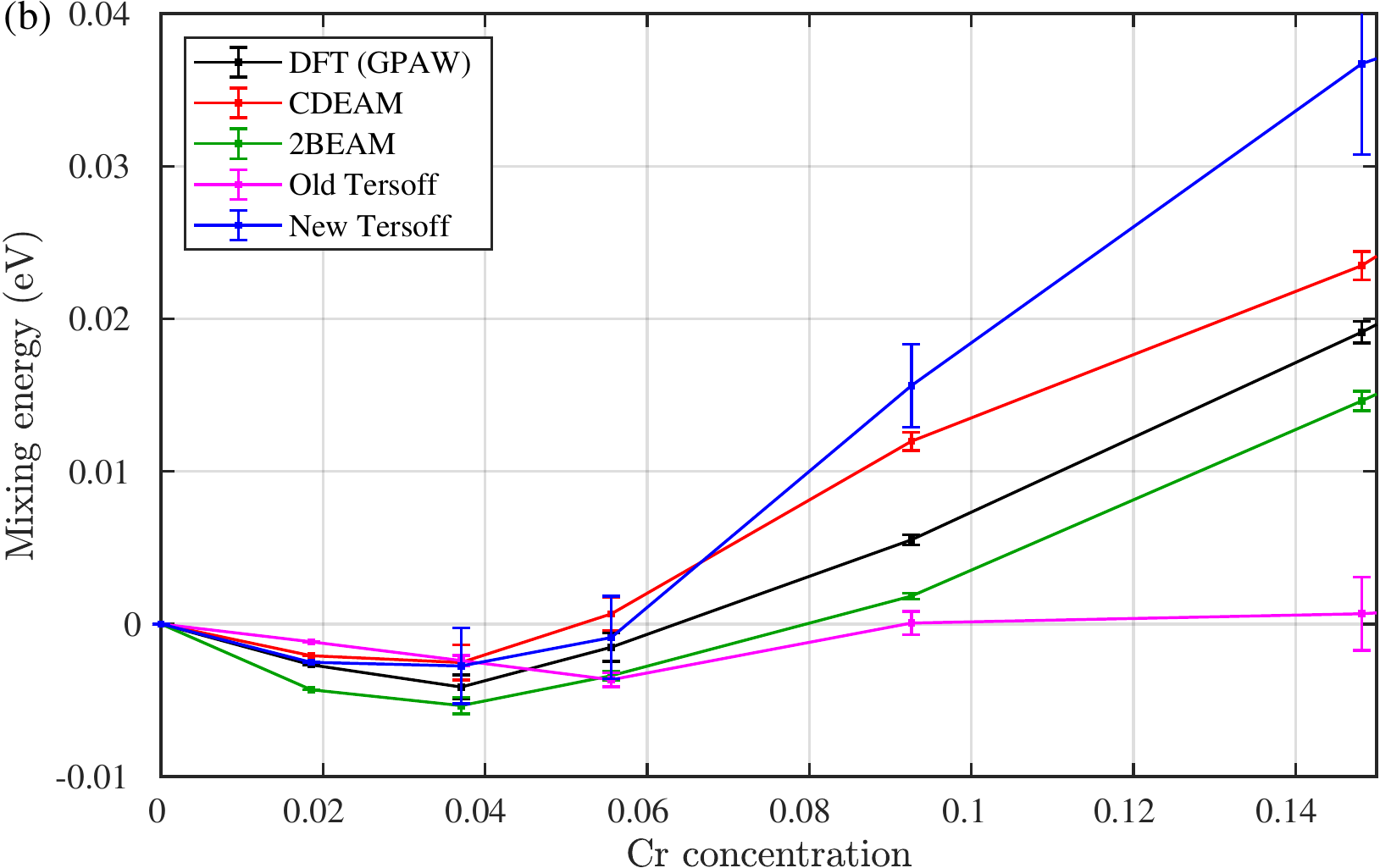}
\caption{\label{fig:mixing_energy}(a)~Mixing energy of the bcc Fe--Cr alloy as a function of the chromium concentration as given by our DFT calculations and the four Fe--Cr potential models. (b)~Close-up of the region below 15\% Cr concentration. The calculations are performed using a 54-atom computational cell. A total of 131 different 54-atom configurations are used for the whole Cr concentration range from 0 to 1, with the error bars corresponding to the standard error of the mean at the given Cr concentration. The lines drawn through the data points are guides to the eye.
\softpekcom{I should say something about the configurations used either here or in the body text.} 
}
\end{figure}

\softpekcom{Start the bulk results with introducing Fig.~\ref{fig:mixing_energy} and defining the mixing energy and formation energy.}

\begin{figure}[b] 
\includegraphics[width=0.99\columnwidth,keepaspectratio]{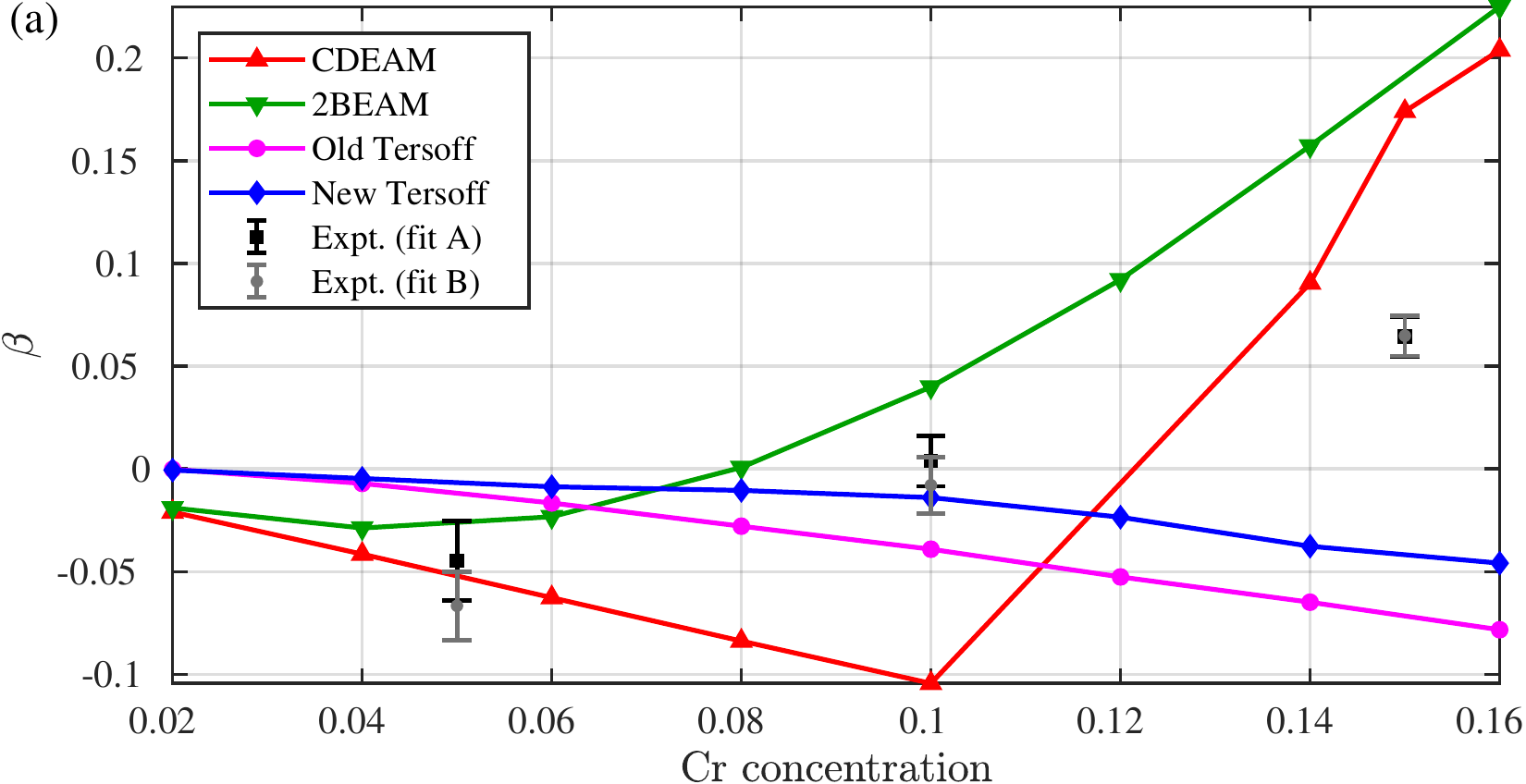}\\
\vspace{4mm}
\includegraphics[width=0.99\columnwidth,keepaspectratio]{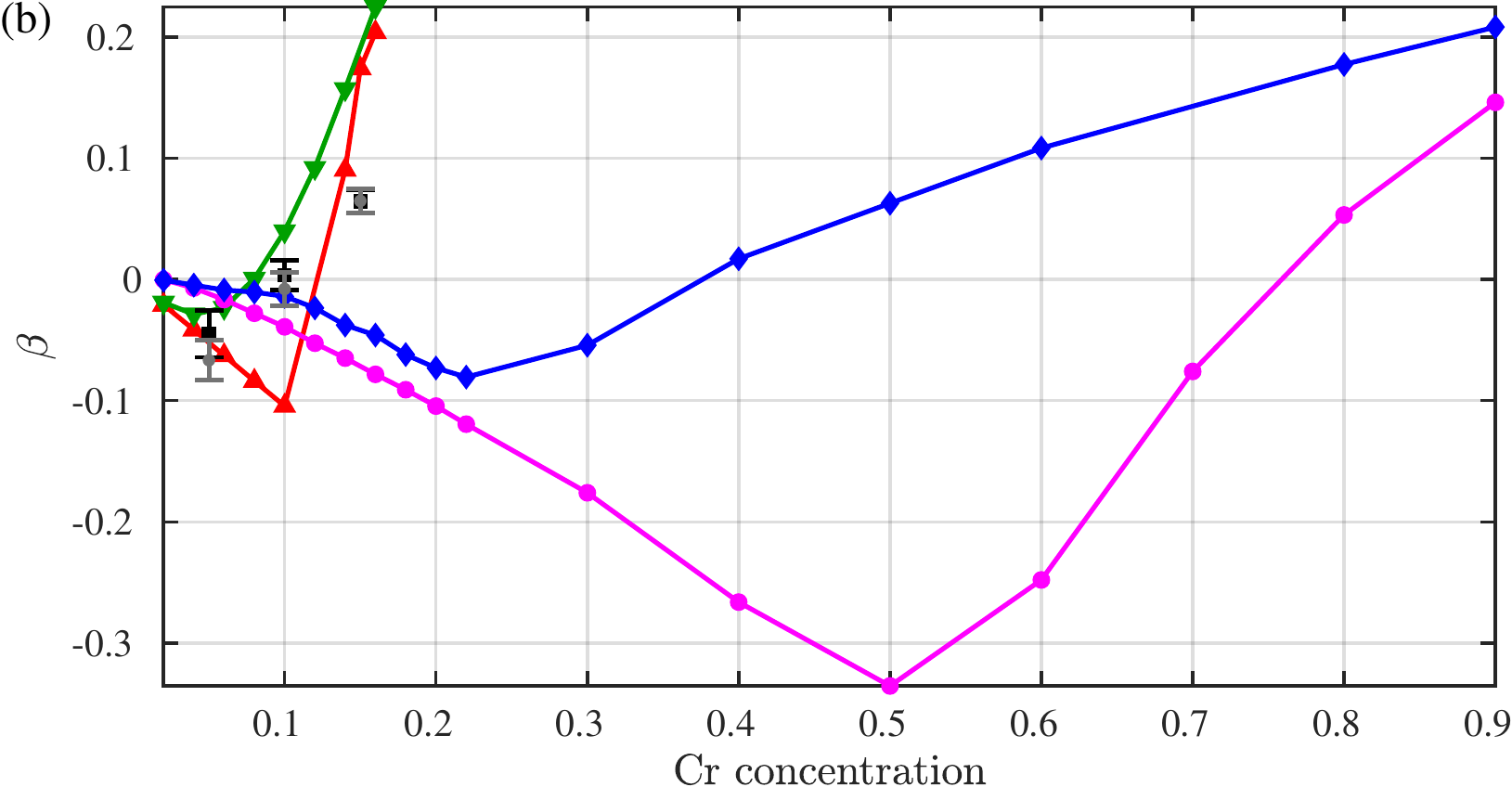}
\caption{\label{fig:sro_700K}Short-range order parameter $\beta$ [Eq.~\eqref{eq:sro_beta}] at \SI{700}{\kelvin} in a bcc Fe--Cr alloy as a function of the chromium concentration $\Crconc$ for (a)~$0.02 \leq \Crconc \leq 0.16$ and (b)~$0.02 \leq \Crconc \leq 0.9$. The lines are guides to the eye. The experimental data points at $\Crconc=0.05$, $0.1$, and $0.15$ are from the diffuse--neutron-scattering measurements of Mirebeau, Hennion, and Parette~\cite{Mir1984.PRL53.687}; fits A and B correspond to least-squares fits of the nuclear cross section with three and four order parameters, respectively.
\softpekcom{Could present these at other temperatures as well, but there is no experimental data to compare against.}
} 
\end{figure}

Let us start with the properties of bulk Fe--Cr alloys and compare the predictions of the new Tersoff potential to those of our DFT calculations, the CDEAM potential, the 2BEAM potential, and the old Tersoff potential. 
In particular, Figs.~\ref{fig:mixing_energy} and~\ref{fig:sro_700K} show, respectively, the mixing energy and the Cowley short-range order parameter of the bcc Fe--Cr alloy as a function of the chromium concentration, and Table~\ref{table:defect_energies} lists the formation energies of various isolated Cr defects in bcc iron.
In Fig.~\ref{fig:mixing_energy}, the alloy mixing energy is determined as the formation energy per atom averaged over different random-alloy configurations 
at a given chromium concentration $\Crconc \coloneqq N_{\Cr}/\left(N_{\Fe}+N_{\Cr}\right)$. 
The formation energy of a given structure, in turn, is defined in this work as
\begin{equation}\label{eq:formation_energy}
\eform = \etot\left(\Fe_{1-\Crconc}\Cr_{\Crconc}\right) - N_{\Fe}\ecoh \left(\Fe\right) - N_{\Cr}\ecoh\left(\Cr\right),
\end{equation}
where $\etot$ is the total potential energy of the computational cell, $N_\elemindex{}$ is the number of atoms of element $\elemindex{}$ in the cell, and $\ecoh\left(\elemindex{}\right)$ is the cohesive energy of element $\elemindex{}$. 
The cohesive energy $\ecoh\left(\elemindex{}\right)$ is determined as the total potential energy of a cell containing only atoms of element $\elemindex{}$ divided by the number of atoms in that cell.
For the Fe--Fe and Cr--Cr potentials in Table~\ref{table:abop_parameters}, 
$\ecoh\left(\Fe\right)=\SI{-4.179}{\electronvolt}$ (bcc lattice constant $a_\Fe=\SI{2.889}{\angstrom}$) and $\ecoh\left(\Cr\right)=\SI{-4.099}{\electronvolt}$ ($a_\Cr=\SI{2.872}{\angstrom}$). 

As reported by Olsson, Abrikosov, Vitos, and Wallenius~\cite{Ols2003.JNuclMater321.84}, \emph{ab initio} calculations predict a negative (positive) mixing energy of Fe--Cr alloys for chromium concentrations below (above) $\sim 6\%$. As can be seen from Fig.~\ref{fig:mixing_energy}(b), our DFT calculations corroborate this result. Although the zero-crossing behavior is qualitatively reproduced by all four potential models under consideration, 
there are quantitative differences: the new Tersoff potential yields the best fit to the \emph{ab initio} mixing energies for Cr concentrations $<6$\%, closely followed by the CDEAM potential, while the 2BEAM and old Tersoff potentials predict the zero crossing to lie at a higher Cr concentration [Fig.~\ref{fig:mixing_energy}(b)]. For chromium-rich alloys, which are included in Fig.~\ref{fig:mixing_energy}(a), the new Tersoff potential matches the DFT values significantly better than the other three potentials.
For example, at 50\% Cr concentration, where the alloy mixing energy has been calculated with the SQS cell~\cite{Zun1990.PRL65.353}, the mixing energy given by the new Tersoff potential is only 0.2\% larger than the DFT value of \SI{4.663}{\electronvolt}, whereas the CDEAM, 2BEAM and old Tersoff potentials differ, respectively, by 15\%, $-9.7$\%, and $-65$\% from the DFT result.
\softpekcom{Let me know if you think I should invert the order of the panels in Fig.~\ref{fig:mixing_energy}. \akcom{Ok for me.}}

\softpekcom{As the second set of bulk data, present the point defect energies of Table \ref{table:defect_energies}. Check also if increasing the computational cell size changes the formation energies appreciably, i.e., whether there is interactions between the defects.}

%

\begin{table*}[tb]
\caption{\label{table:defect_energies}Energies of Cr point defects in bcc Fe, as per our DFT calculations and the CDEAM, 2BEAM, the old Tersoff, and the new Tersoff potentials. For consistency with the DFT treatment, all the formation energies are calculated using a 54-atom cell. The last line is for the Fe substitutional defect in bcc Cr.}
\begin{ruledtabular}
\begin{tabular}{lddddd}
& \multicolumn{5}{c}{Formation energy (eV)} \\
\cline{2-6}
\multicolumn{1}{c}{Defect} & \multicolumn{1}{c}{DFT (GPAW)}&  \multicolumn{1}{c}{CDEAM} & \multicolumn{1}{c}{2BEAM} & \multicolumn{1}{c}{Old Tersoff} & \multicolumn{1}{c}{New Tersoff} \\
\hline
$\langle 100 \rangle$ Fe--Cr & 5.327 & 3.56 & 3.66 &4.96 &5.15  \\
$\langle 110 \rangle$ Fe--Cr & 4.090 & 3.20 & 3.18 & 4.06 &4.22 \\
$\langle 111 \rangle$ Fe--Cr & 4.596 & 3.19 & 3.54 & 4.66 &4.88 \\
Octahedral Cr & 5.339 & 3.56 & 3.33 & 6.01 & 5.30 \\
Tetrahderal Cr & 4.611 & 3.50 & 3.32 & 4.44 & 4.83\\
Substitutional Cr & -0.1444 & -0.112 & -0.233 & -0.0629  & -0.136\\
$\langle 100 \rangle$ Cr--Cr & 5.43 &4.49  & 3.98 & 4.88 & 5.45 \\
$\langle 110 \rangle$ Cr--Cr & 4.35 & 3.81 & 3.26 & 3.80 & 4.09\\
$\langle 111 \rangle$ Cr--Cr & 4.699 & 4.63 & 3.51 & 4.58 & 5.00 \\
\hline
Substitutional Fe in Cr & 0.4825 & 0.498 & 0.357 & 0.235  & 0.511
\end{tabular}
\end{ruledtabular}
\end{table*}

The formation energies of selected Cr point defects in bcc Fe are presented in Table~\ref{table:defect_energies}. In general, all four potential models are in fairly good agreement with the target DFT values, with percentage errors typically $<10$\%. If we take the nine Cr point defects listed in Table~\ref{table:defect_energies} and compute the symmetric mean absolute percentage error (SMAPE) of the predictions of each potential model~\cite{[{The SMAPE is defined here as 
$100\%\sum_{j=1}^{N_E} \abs{t_j-p_j}/\bigl(N_E \abs{t_j}+N_E \abs{p_j}\bigr)$, where the sum is over the $N_E=9$ formation energies and $t_j$ and $p_j$ are their target values and potential-model predictions, respectively. See, e.g., }]Arm1985.book,*Flo1986.Omega14.93},
we obtain the following ranking (from best to worst): new Tersoff (SMAPE of 2.1\%), old Tersoff (7.2\%), CDEAM (13\%), and 2BEAM (17\%). On average, the new Tersoff potential is therefore in a better agreement with our \emph{ab initio} point-defect energies than the other three potential models. We note in particular that the new Tersoff potential provides the best match for the negative DFT value of the Cr substitutional defect.

\softpekcom{Next, define the SRO parameter and present and analyze the SRO data (at three different temperatures?). In particular, compare the various model predictions with the experimental data at \SI{700}{\kelvin} (Fig.~\ref{fig:sro_700K}).} 

Besides comparing formation energies, we may examine the degree of short-range ordering in a thermally equilibrated Fe--Cr alloy. To this end, the Cowley short-range order parameters $\alpha$ can be defined as~\cite{Cow1950.PR77.669}
\begin{equation}
\alpha_{\Cr}^{(k)} = 1 - 
\frac{Z^{(k)}_{\Fe}}{\left(Z^{(k)}_{\Fe}+Z^{(k)}_{\Cr}\right)\left(1-c_{\Cr}\right)},
\end{equation}
where $Z^{(k)}_{\Fe}$ and $Z^{(k)}_{\Cr}$ are the average numbers of Fe and Cr atoms in the $k$th neighbor shell.
We further define the linear combination
\begin{equation}\label{eq:sro_beta}
\beta=\frac{8 \alpha_{\Cr}^{(1)} + 6\alpha_{\Cr}^{(2)}}{14}
\end{equation}
relevant to bcc lattices. If $\beta < 0$, a Cr atom prefers to have Fe atoms as its nearest neighbors; if $\beta > 0$, Cr prefers Cr neighbors; $\beta=0$ corresponds to a random alloy. The configurations used to determine $\beta$ are computed with a Monte Carlo method where possible moves 
consist of atom displacements and exchanges of types (Fe or Cr) of pairs of atoms.
The displacements are performed with short sequences of MD simulations in the canonical ensemble, 
because this 
has been found to be more efficient in moving the atoms than the conventional Metropolis algorithm~\cite{Kur2015.PRB92.214113}. It should be noted that these Monte Carlo--MD calculations are pure equilibrium simulations with no kinetics involved.

Figure~\ref{fig:sro_700K} shows the order parameter $\beta$ at \SI{700}{\kelvin} as a function of the Cr concentration 
for the four potential models, along with three experimental data points determined by 
diffuse--neutron-scattering measurements~\cite{Mir1984.PRL53.687}.
While none of the four potential models yields a particularly good fit to the experimental data, the CDEAM and 2BEAM potentials at least qualitatively reproduce the observed change of sign of $\beta$ from negative to positive around $\Crconc = 0.1$  [Fig.~\ref{fig:sro_700K}(a)].
For both Tersoff potentials, $\beta$ remains negative up to high Cr concentrations, eventually turning positive at $\Crconc\approx 0.37$ for the new Tersoff potential and at $\Crconc\approx 0.76$ for the old Tersoff potential [Fig.~\ref{fig:sro_700K}(b)].
We thus conclude that the new Tersoff potential, although largely failing to match the experimental data, still performs noticeably better than the old Tersoff potential in describing the short-range ordering in Fe--Cr alloys.

\begin{figure}[htb]
\includegraphics[width=0.65\columnwidth,keepaspectratio]{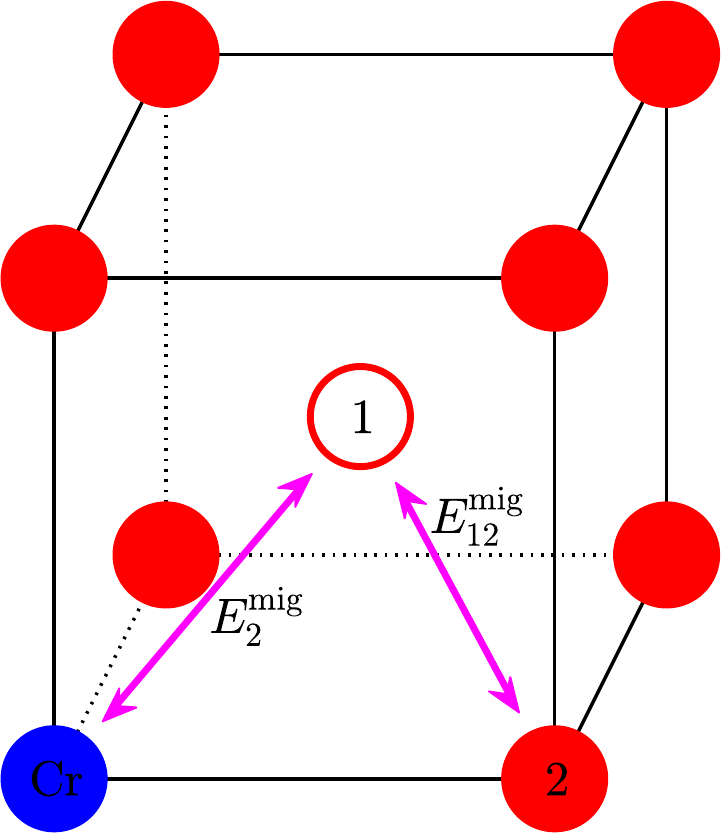}
\caption{\label{fig:energy_barriers}Nomenclature used for the migration energy barriers. The Cr atom is shown in blue and the Fe atoms in red. $E_2^\mathrm{mig}$ denotes the migration energy barrier for the solute--vacancy jump, and $E_{ij}^\mathrm{mig}$ is for an iron atom moving from site $j$ to a vacant site $i$. Table~\ref{table:migration_barriers} lists the values of these energies barriers within the different models under consideration. 
}
\end{figure}

\softpekcom{As the final set of bulk results, present migration energy barriers in bulk bcc Fe (Fig.~\ref{fig:energy_barriers} and Table~\ref{table:migration_barriers}).}

We have also computed the migration energy barries related to the diffusion of a vacancy--Cr-substitutional pair in bulk bcc Fe using the nudged elastic band (NEB) method~\cite{Mil1995.SurfaceScience324.305,Jon1998.chapter} implemented in LAMMPS. We consider both a process where a Cr subsitutional moves to a vacancy in the nearest-neighbor site and a process where the vacancy migrates from the nearest-neighbor to the second-nearest-neighbor site of the Cr atom (see Fig.~\ref{fig:energy_barriers}). The obtained barrier energies for the four potential models are listed in Table~\ref{table:migration_barriers} along with DFT results by Messina, Nastar, Garnier, Domain, and Olsson~\cite{Mes2014.PRB90.104203}.
By far the best agreement with the DFT values is given by the 2BEAM potential (with the largest relative difference being $<6\%$), followed by the CDEAM potential, which overestimates the Cr--vacancy migration energy by 53\% but is otherwise close to the target values. The new Tersoff potential is in fairly good agreement with DFT for the Cr--vacancy migration energy $E_2^\mathrm{mig}$ but noticeably overestimates both barries $E_{12}^\mathrm{mig}$ and $E_{21}^\mathrm{mig}$. The old Tersoff potential is also off by significant margins (26\% for $E_2^\mathrm{mig}$, $-32\%$ for $E_{12}^\mathrm{mig}$, and $16\%$ for $E_{21}^\mathrm{mig}$). 

\begin{table}[tbh]
\caption{\label{table:migration_barriers}Migration energy barriers for Cr--vacancy diffusion and Fe self-diffusion in the neighboring site (see Fig.~\ref{fig:energy_barriers}).}
\begin{ruledtabular}
\begin{tabular}{lddd}
Method & E_2^\mathrm{mig} &  E_{12}^\mathrm{mig}   & E_{21}^\mathrm{mig}  \\
\cline{1-4}
DFT~\cite{Mes2014.PRB90.104203} & 0.575(45) & 0.69 & 0.64(1) \\
CDEAM & 0.8772 & 0.6467 & 0.6160 \\
2BEAM & 0.5725 & 0.6511 & 0.6260 \\
Old Tersoff & 0.7267 & 0.4666 & 0.7413 \\
New Tersoff & 0.632 & 1.09 & 0.941 
\end{tabular}
\end{ruledtabular}
\end{table}

\subsection{Comparison of the potential models for surface-related properties}\label{subsec:surface_results}

We now move to scenarios involving surfaces 
of Fe--Cr alloys and investigate whether we have achieved our goal of improving 
upon the performance of the existing potential models in 
predicting properties of such surfaces. 

Let us first consider the segregation of Cr atoms to the (100) surface of bcc Fe. The segregation energy of Cr from a given reference region A to another region B is defined as 
the net change in energy when a Cr atom is transferred from A to B and an Fe atom from B to A,
\begin{equation}
E^{\mathrm{Cr}}_{\mathrm{segr},\mathrm{A}\to\mathrm{B}}=\etot(\textrm{Fe at A, Cr at B})-\etot(\textrm{Fe at B, Cr at A}),
\end{equation}
where $\etot$ is the total energy of the computational cell.  In our case, B will be one of the top surface layers and A will be a bulk site. The energies are calculated using a slab geometry with periodic boundary conditions in $y$ and $z$ directions and the slab thickness $L_x$ large enough that further increases in $L_x$ cause no changes in the segregation energies. The reference bulk position A is taken from the middle of the slab. For nonzero background Cr concentration, the segregation energy values are averaged over 10\,000 random alloy configurations.

\begin{figure}[htb] 
\includegraphics[width=0.891\columnwidth,keepaspectratio]{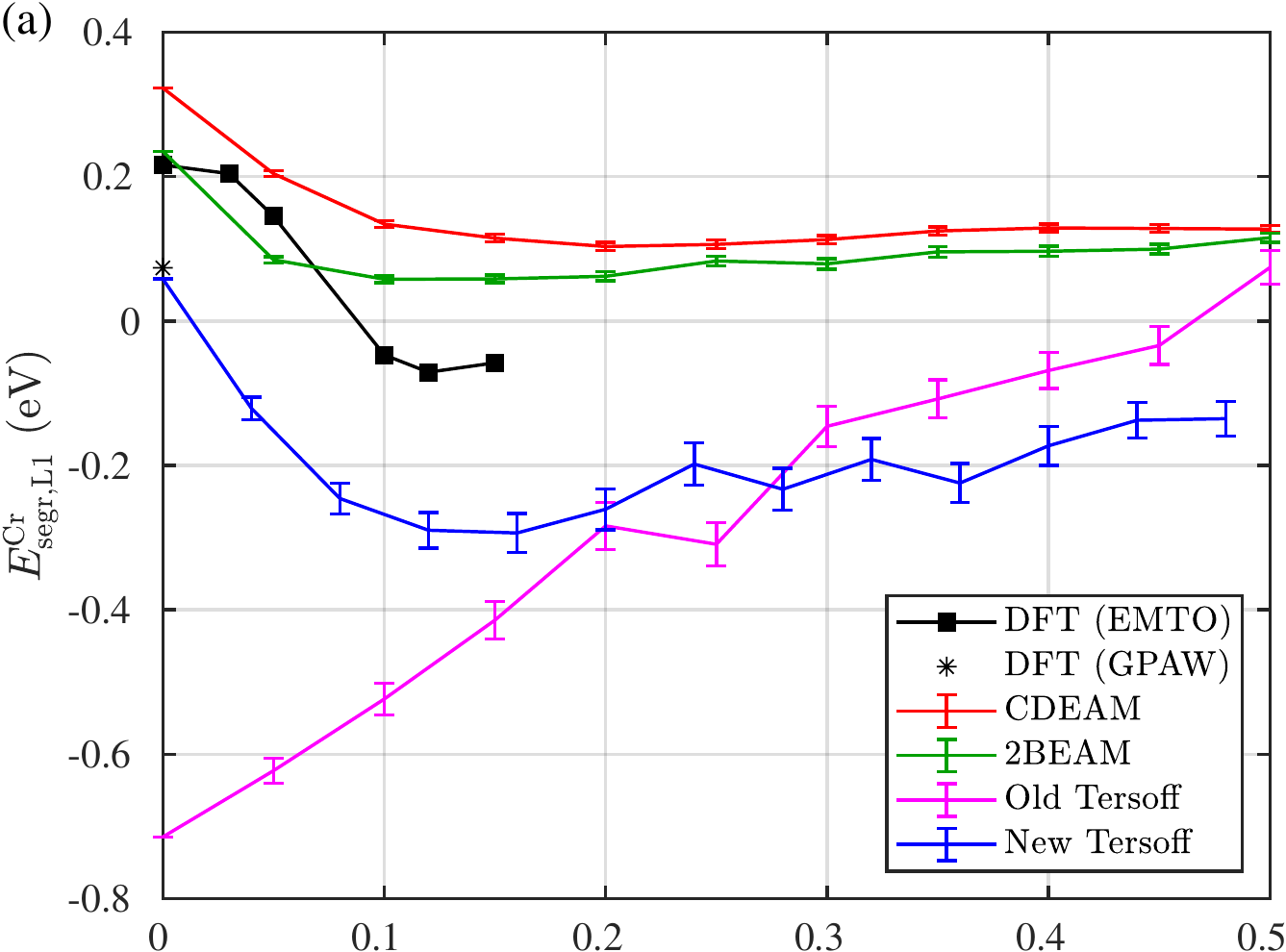}
\includegraphics[width=0.891\columnwidth,keepaspectratio]{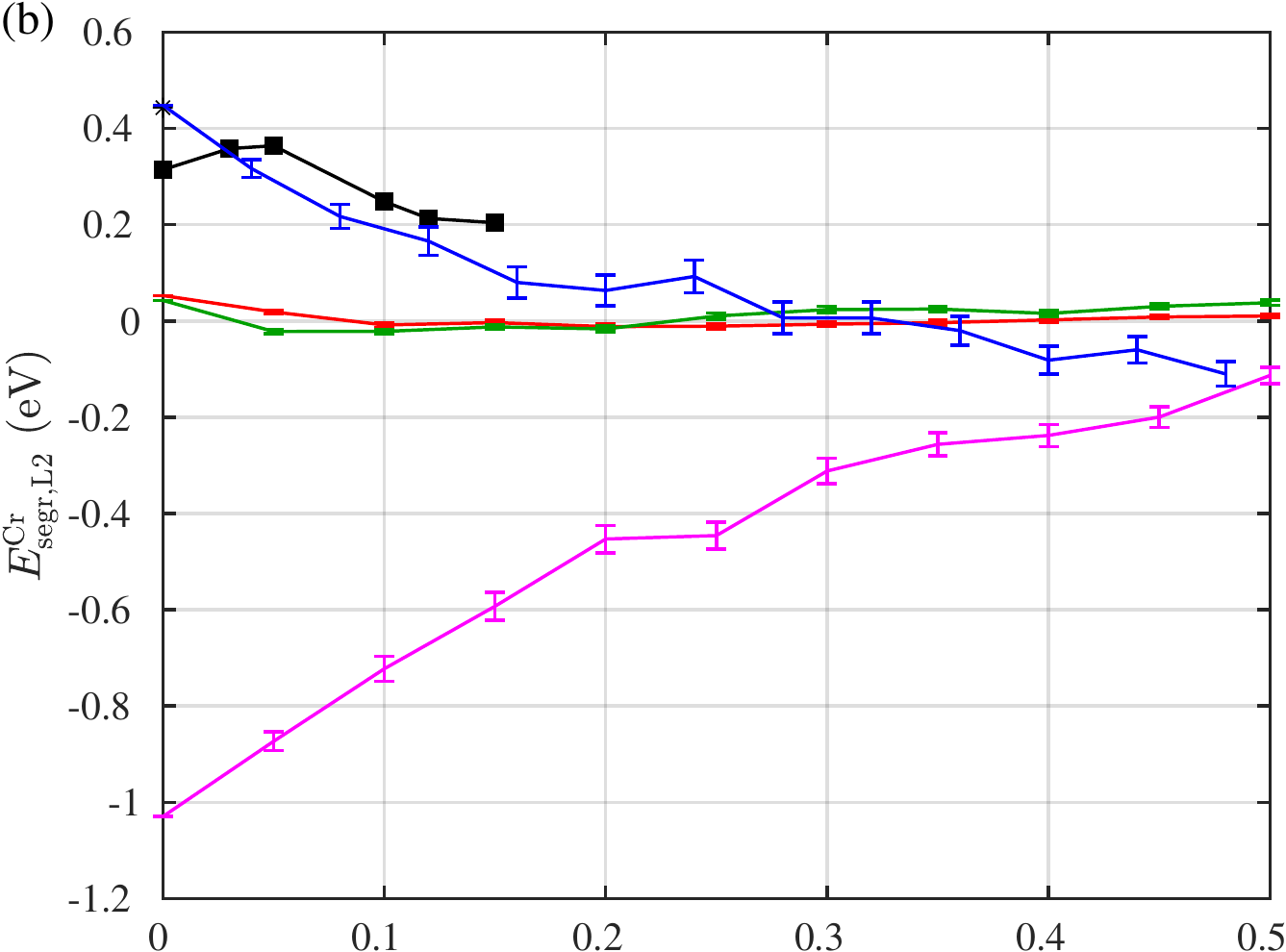}
\includegraphics[width=0.891\columnwidth,keepaspectratio]{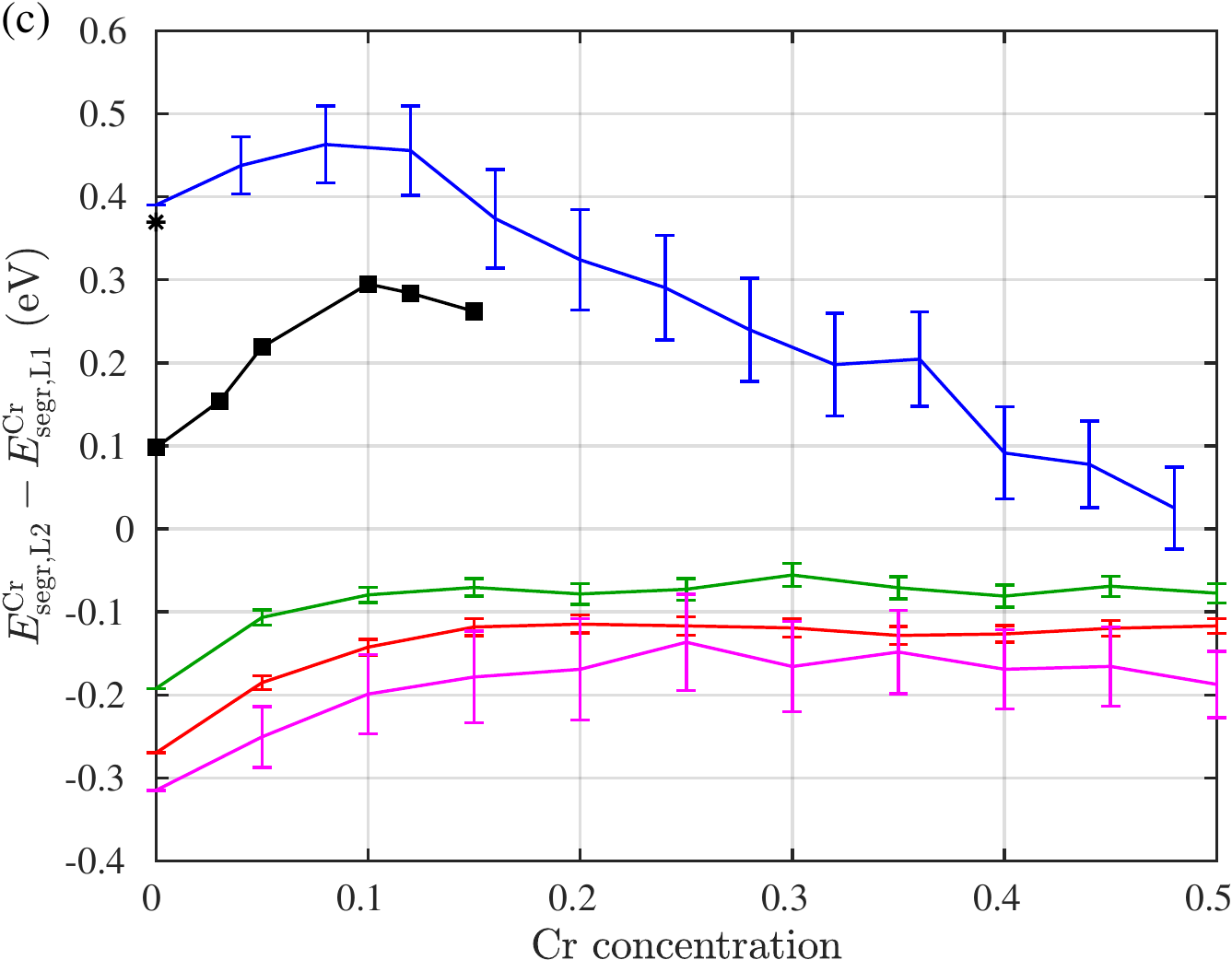}
\caption{\label{fig:segregation_energy}Segregation energy of a Cr atom in a bcc Fe--Cr alloy from the bulk to (a) the surface layer and (b) the first subsurface layer as a function of the chromium concentration $\Crconc$. (c) The difference between the surface- and first-subsurface-layer segregation energies. The asterisk is the GPAW \emph{ab initio} result at $\Crconc=0$ and the squares are the \emph{ab initio} results from Ref.~\cite{Kur2015.PRB92.214113} obtained using the EMTO method. The lines are guides to the eye.}
\end{figure}

The Cr segregation energies to the topmost surface layer ($\eseg{1}$) and the second topmost layer ($\eseg{2}$) are shown in Figs.~\ref{fig:segregation_energy}(a) and ~\ref{fig:segregation_energy}(b), respectively, as a function of the Cr concentration $\Crconc$ for both the DFT and the four potential models. Due to computational limitations, the GPAW DFT calculations are limited to zero background Cr concentration. The DFT results for $\Crconc > 0$ are obtained 
with a basis set of exact muffin-tin orbitals (EMTO)~\cite{And1994.incollection,And2000.PRB62.R16219,Vit2007.book} in combination with the coherent potential approximation~\cite{Sov1967.PR156.809,Vit2001.PRL87.156401}, 
which circumvents the need to average over a large number of alloy configurations as done for the potential models. 
Note that only the zero-concentration segregation energies of the new Tersoff potential have been explicitly fitted (with the target values being the GPAW ones); the data for $\Crconc > 0$ should be regarded as predictions of the model.

The old Tersoff potential from Ref.~\cite{Hen2013.JPCM25.445401} is observed to be far away from the \emph{ab initio} segregation energies for both layers.
The CDEAM and 2BEAM potentials give a fairly good fit to the EMTO results for $\eseg{1}$ but are far off from both the GPAW and EMTO results for $\eseg{2}$. 
Importantly, as can be observed from Fig.~\ref{fig:segregation_energy}(c), all three pre-existing models (CDEAM, 2BEAM, and old Tersoff) yield $\eseg{2}-\eseg{1} < 0$ at all Cr concentrations shown and thus predict segregation of Cr to the second atomic layer instead of the topmost one. This is in stark contrast to the \emph{ab initio} results, for which  $\eseg{2}-\eseg{1}$ is positive at all investigated concentrations. Fortunately, this shortcoming is fixed by our new Tersoff potential, for which $\eseg{2}-\eseg{1} > 0$ in the entire range $0 \leq \Crconc \leq 0.48$ and the fit to the GPAW value of  $\eseg{2}-\eseg{1}=\SI{0.3696}{\electronvolt}$ at $\Crconc=0$ is excellent, the relative error being $2.5\%$.

We have also investigated the migration of Cr in the vicinity of an Fe(100) surface. 
In Table~\ref{table:surface_migration_barriers}, we list the migration energy barriers for the process where a surface Cr atom moves to a vacancy at a nearest-neighbor site in the second layer. Although we do not have a DFT value for this migration energy to compare against, it is worthwhile to note that the 2BEAM potential and the new Tersoff potential both yield 
an energy barrier close to zero 
whereas the value given by the old Tersoff potential is much higher than the others.

\begin{table}[tbh]
\caption{\label{table:surface_migration_barriers}Migration energy barriers for a surface Cr atom moving to a nearest-neighbor second-layer vacancy.}
\begin{ruledtabular}
\begin{tabular}{ld}
Potential & \multicolumn{1}{c}{Energy barrier (eV)}  \\
\cline{1-2}
CDEAM & 0.2624 \\
2BEAM & 0.0035 \\
Old Tersoff & 1.1866 \\
New Tersoff & 0.0023 
\end{tabular}
\end{ruledtabular}
\end{table}

\softpekcom{Here we can also present the migration energy barriers as a function of the layer index (cf. Daniel's presentation). No DFT data to compare against, though.}

\section{Discussion}\label{sec:discussion}

In summary, we have presented a new interatomic Fe--Cr potential that is suitable for molecular dynamics simulations of surface physics in Fe--Cr alloys. The potential was formulated in the Tersoff formalism, allowing it to be combined with previously developed Fe--C and Cr--C potentials~\cite{Hen2013.JPCM25.445401} into a model of the stainless-steel system Fe--Cr--C. The new potential parameters were optimized by fitting to a structural database consisting of \emph{ab initio} results not only for bulk alloys but also for the segregation of Cr atoms 
to the (100) surface of an Fe--Cr alloy. 
We compared the performance of the new potential to that of three pre-existing Fe--Cr potentials with regard to the fitting database as well as to bulk- and surface-related quantities not involved in the fitting.

For the bulk properties, the new Tersoff potential was found to perform at the same overall level as the pre-existing EAM potentials considered, namely, the CDEAM and the 2BEAM potentials. On the one hand, the new Tersoff potential provided the closest match of all the tested potentials to our DFT-calculated mixing energies and point-defect energies. On the other hand, the 2BEAM potential was the best performer when it came to the short-range ordering in random Fe--Cr alloys 
(qualitatively reproducing the experimentally observed sign change of the order parameter at $\Crconc \approx 0.1$)
and to the vacancy--Cr diffusion in bcc Fe 
(yielding energy barriers within 6\% of the DFT values). 
For the tested bulk data, the new Tersoff potential performed significantly better than the old Tersoff potential from Ref.~\cite{Hen2013.JPCM25.445401}, even though only the latter was fitted exclusively to bulk quantities.

The main objective we set at the beginning was
for the new potential model
to outperform 
the existing models 
by correctly predicting the surface-segregation behavior of Cr in Fe--Cr alloys.

We achieved this goal in the sense that the new Tersoff potential yields the same ordering of Cr segregation energies as the DFT calculations, namely, $\eseg{1} < \eseg{2}$. This is in contrast to the other potential models, for which $\eseg{1} > \eseg{2}$ 
and which thus predict Cr to segregate primarily to the second layer instead of the first.
The agreement between the new Tersoff potential and the DFT calculations is far from perfect, however. For example, the new Tersoff potential predicts 
a sign change in the first-layer segregation energy
at a significantly smaller Cr concentration ($\sim 2.5\%$) than do the DFT-based EMTO calculations ($\sim 8\%$)~\cite{Kur2015.PRB92.214113,Rop2007.PRB76.220401}.

In light of the above, we conclude that the new Tersoff potential appears to be the best potential for scenarios where both surface and bulk properties of Fe--Cr alloys are of importance. In bulk systems without boundaries, however, the 2BEAM potential is perhaps the most optimal choice, especially if we take into account its lower computational cost compared to the Tersoff potentials. The old Tersoff potential performs worse than the new one in almost all of our tests, and hence 
it seems difficult to justify its future use.

\begin{acknowledgments}
This work was supported by the Academy of Finland (Grant No.~308632). The computational resources granted by the CSC -- IT Center for Science, Finland, and by the Finnish Grid and Cloud Infrastructure project (FGCI; urn:nbn:fi:research-infras-2016072533) are gratefully acknowledged, as are the facilities provided by the Turku University Centre for Materials and Surfaces (MatSurf).
\end{acknowledgments}
\bibliographystyle{apsrev4-1}
\bibliography{pak_bib_amksia,gpaw_ref}
\end{document}